\documentclass[12pt]{article}
\usepackage{epsfig,amsfonts,amssymb}
\usepackage{amsmath}
\usepackage[margin=2.5cm]{geometry}
\usepackage[pdftex, breaklinks=true, linktocpage,
colorlinks=true, urlcolor=blue, linkcolor=blue, citecolor=red
]{hyperref}
\usepackage[all]{hypcap}
\usepackage[colorlinks=true,urlcolor=blue,linkcolor=blue]{hyperref}
\usepackage{cite}
\usepackage{hyperref}
\usepackage{framed}
\usepackage{breqn}
\usepackage{flexisym}
\usepackage[english]{babel}

\usepackage{ifpdf}
\usepackage[utf8]{inputenc}
\usepackage{longtable}
\usepackage{young}
\usepackage[makeroom]{cancel}
\usepackage[normalem]{ulem}
\usepackage{framed}
\usepackage{verbatim}
\usepackage{fancyhdr}
\usepackage[all]{hypcap}
\usepackage{bm, bbm}
\usepackage[usenames, dvipsnames]{xcolor}

\usepackage{eurosym, xspace}
\usepackage[nottoc]{tocbibind}
\usepackage{graphicx, subfig, float}
\usepackage[multiple]{footmisc}
\usepackage{authblk}

\graphicspath{{images/}}
\usepackage[sort&compress, english]{cleveref}

\input epsf.sty
\newcommand{\refb}[1]{(\ref{#1})}

\newcommand{\Pf}{\mathop{\mathrm{Pf}}}

\newcommand{\sgn}{\mathop{\mathrm{sgn}}}
\begin{document}

\baselineskip 24pt

\vskip .6cm
\medskip

\vspace*{4.0ex}

\baselineskip=18pt
\begin{center}
	{\Large \bf Super-maximal chaos and instability}

	\vskip .6cm
	\medskip
	
	\vspace*{4.0ex}
	
	\baselineskip=18pt
	{\large \rm Swapnamay Mondal}
	\vspace*{4.0ex}
	
	\textit{
		\vskip0.2cm
		Sorbonne Universit\'es, Paris 06,  \\ 
		UMR 7589, LPTHE, 75005, Paris, France} \\
		\vskip0.2cm
	\vskip0.5cm
	
	\vspace*{1.0ex}
	\small swapno@lpthe.jussieu.fr
\end{center}
\vspace*{4.0ex}

\vspace*{1.0ex}

\vspace*{5.0ex}

\centerline{\bf Abstract} \bigskip
An upper bound on Lyapunov exponent of a thermal many body quantum system has been conjectured recently. In this work, we attempt to achieve a physical understanding of what prevents a system from violating this bound. To this end, we propose - super-maximal chaos leads to instability. Our proposal is supported by findings in a SYK lattice model, with a tuneable parameter, which the Lyapunov spectrum depends upon. In the stable regime of this parameter, along with incoherent metallic phase, the system exhibits another novel phase, where transport is controlled neither by quasi-particles nor by diffusion. At the phase transition, diffusion coefficient, butterfly velocity and Thouless time diverges.


\vfill \eject

\baselineskip=18pt

\tableofcontents

\section{Introduction}  \label{s1}
It has been recently conjectured \cite{Maldacena:2015waa} that in a many body quantum system, with a large hierarchy between scrambling time and dissipation time, chaos can develop no faster than in Einstein gravity. This amounts to a bound on Lyapunov exponent $\lambda_L \leq 2\pi T$, where $T$ is the temperature\footnote{For systems without quasi particles, in general one expects $\lambda_L \sim T$, unlike systems with quasi-particles where one usually has $\lambda_L \sim T^a$ with $a>1$.}. The conjecture is motivated by study of several examples as well as analysis of mathematical properties of out of time order correlation functions.

Since Einstein gravity saturate the chaos bound, a quantum system with maximal chaos is likely to have a dual gravity description. This insight has been crucial in current investigations in black hole physics. Quantum mechanical models exhibiting maximal chaos (among other curious features) has been proposed as models for near extremal black holes \cite{K2}, \cite{Witten:2016iux}. Also see \cite{Maldacena:2016hyu,Polchinski:2016xgd,Berkooz:2016cvq, Gross:2016kjj,Klebanov:2016xxf,Peng:2016mxj,Turiaci:2017zwd,Fu:2016vas}. Bulk dual of these models have been discussed in \cite{Gross:2017hcz,Gross:2017aos,Das:2017pif,Das:2017hrt,Das:2017wae} . This is not the topic of this paper though. We simply attempt to understand the chaos bound a bit better.

A universal bound in Lyapunov exponent is rather intriguing, since such a bound is not part of basic principles of physics. For example, this is qualitatively different from having an upper limit on velocity of a physical object. Perhaps a way to understand this bound better would be to ask- what goes wrong if a system violates this bound? In this work we suggest the answer to this question could be that the system becomes unstable. To put colloquially, too much chaos leads to instability. 

Our suggestion is based on our findings in a SYK lattice model with a tuneable locality parameter $\alpha$. When $\alpha$ crosses a $1/2$ from right, the leading Lyapunov exponent violates the chaos bound! On closer inspection, one discovers that the system becomes unstable at the same point, therefore rendering the analysis of chaos invalid and salvaging the chaos bound of \cite{Maldacena:2015waa}.

Apart from providing insights about chaos bound, this system also exhibits interesting phase structure in $1/2 < \alpha < \infty$ range. For $1< \alpha < \infty$, the system behaves as a diffusive incoherent metal. At $\alpha \rightarrow 1_+$, the diffusion coefficient diverges and so does butterfly velocity and Thouless time (after appropriate regularisation). The divergence of diffusion coefficient signals manifestation of non-locality and that of Thouless time signals absence of ergodic phase \cite{Altland:2017eao}. For $\alpha =1$, the diffusion coefficient is logarithmically renormalised . As one moves towards even smaller values of $\alpha$, this renormalisation becomes even stronger leading to qualitative change in the form of density density correlators. Diffusion coefficient does not seem to be an useful concept in this regime.

The paper is organised as follows. In \ref{sdefn}, we introduce our model. Two point and four point functions are analysed in \ref{s2pt} and \ref{s4pt} respectively. Chaos and apparent violation of chaos bound is discussed in \ref{schaos}. In \ref{sstability} we analyse stability of the mean field solution. Transport properties and phase structure of the model is discussed in \ref{sphase}. Finally we summarise our work and discuss future directions in \ref{sdscsn}.

\section{The Model} \label{s2}
The maximally chaotic non-Fermi liquid phase of SYK model does not persist when perturbed by various interactions \cite{Bi:2017yvx,Chen:2017dav,Banerjee:2016ncu}. It is interesting to ask if one can perturb the SYK model such that the system remains in a chaotic phase, but the Lyapunov exponent depends on the interaction strength and at some critical value violates (or at least attempts to do so) the chaos bound \cite{Maldacena:2015waa}. It turns out that SYK lattice models are useful for posing this question. Such models\footnote{Also see \cite{Chaudhuri:2017vrv,Narayan:2017qtw} for for SYK like tensor models on lattice.} have been studied in \cite{Gu:2016oyy,Jian:2017unn,Jian:2017jfl}. In these models each site hosts a SYK system and then different sites interact in a local fashion. Our model differs from these models in the fact that it is non local in lattice indices\footnote{This is somewhat similar to original SYK model, where Majorana indices were to be thought of as lattice indices.}.
\subsection{Definition of the model} \label{sdefn}
Our model is specified by the Hamiltonian 
\begin{align}
H &= \sum_{x \in \mathbb{Z}} \left[ \sum_{1\leq j <k < l< m \leq N}J_{x;jklm} \chi^x_j \chi^x_k \chi^x_l \chi^x_m + \sum_{y \in \mathbb{Z};\atop y  > x;}\sum_{1\leq j <k \leq N; \atop 1 \leq< l< m \leq N}J'_{(x,y);jklm} |x-y|^{-\alpha} \chi^x_j \chi^x_k \chi^y_l \chi^y_m \,  \right] \, ,
\end{align}
defined on an infinite chain\footnote{The model can be thought of as decompactification limit of a model defined on a circle. This is discussed in \refb{reg}.}. Each site hosts $N$ Majorana fermions, where $N$ is a large number. $\chi^x_i$ denotes $i^{th}$ fermion living in $x^{th}$ site. The random couplings $J$-s are chosen independently for each site, as indicated by the subscript $x$. Similarly $J'$-s are chosen independently for each pair of sites, as indicated by the subscript $(xy)$. The disorder average is specified by
\begin{align}
\overline{J}_{x;jklm} &= 0; ~\overline{J'}_{(xy);jklm} = 0;~\overline{J^2}_{x;jklm} = \frac{6J_0^2}{N^3}; ~\overline{(J')^2}_{(xy);jklm} = \frac{\tilde{J}^2}{N^3} \, .
\end{align}
The locality parameter $\alpha$ is a real number. In $\alpha \rightarrow \infty$ limit, the Hamiltonian reduces to the model studied in \cite{Gu:2016oyy}, containing only nearest neighbour interactions. At any finite value of $\alpha$, the model is non local\footnote{In \cite{Gu:2016oyy} certain non local interactions have been discussed, but not of the kind considered in the present paper.}. The qualitative features of the local limit however persist even deep inside the non local regime, until $\alpha \rightarrow 1_+$, where effects of non locality manifests itself and the system undergoes a phase transition. This is discussed this in \ref{sphase}.

At each site, there is a $\mathbb{Z}_2$ symmetry acting on the fermions as $\chi^x_i \rightarrow - \chi^x_i$. This sets two point functions of fermions of two different sites to zero.
So one has to consider $\langle T \chi^x_i(t_1) \chi^x_i(t_2) \rangle$ only. After disorder average there is an $SO(N)$ symmetry, which further states that  
$\langle T \chi^x_i(t_1) \chi^x_i(t_2) \rangle$ is independent of the flavour index $i$. Same symmetries entail $\langle T \chi^x_i(t_1) \chi^x_i(t_2) \chi^y_j(t_3) \chi^y_j(t_4) \rangle$ is the only non-zero four point function.

 We use the disorder averaged partition function to study this model. Diagrammatic approach should also reproduce the same results. Most computations for finite $\alpha$ closely parallel those for $\alpha \rightarrow \infty$ case, studied in \cite{Gu:2016oyy}.

The disorder averaged partition function is given by
\begin{align}
\nonumber
\overline{Z} &= \int D\chi e^{-N S_{eff}} \, ,\\
\nonumber
\text{where,~} S_{eff} &= \sum_{x \in \mathbb{Z}} \Bigg[ \frac{1}{2} \int d\tau \sum_{i} \chi^x_i \dot{\chi}^x_i - \frac{1}{8N^3} \int_0^\beta \int_0\beta d\tau_1 d\tau_2 \Bigg\{ J_0^2 \left( \sum_{i} \chi^x_i(\tau_1) \chi^x_i(\tau_2)\right)^4 \\
&+ \tilde{J}^2 \sum_{s > 0} s^{-2\alpha} \left( \sum_{i} \chi^x_i(\tau_1) \chi^x_i(\tau_2)\right)^2  \left( \sum_{i} \chi^{x+s}_i(\tau_1) \chi^{x+s}_i(\tau_2)\right)^4  \Bigg\} \Bigg] \, .
\label{partitionfn}
\end{align}
We define the bilocal fields
\begin{align}
G^x(\tau_1, \tau_2) &= \frac{1}{N} \sum_i \chi^x_i(\tau_1) \chi^x_i(\tau_2) \, . \label{bilocal}
\end{align}
To impose \refb{bilocal}, one inserts a delta function in the path integral \refb{partitionfn}, which again can be written as an integral over a new field $\Sigma^x(\tau_1,\tau_2)$. Then one integrates the original Majoranas out to get 
\begin{align}
\nonumber
\bar{Z} &= \int DG D\Sigma~ e^{-NS_{eff}[G,\Sigma]} \\
\nonumber
\text{where,}~~~S_{eff}[G,\Sigma] &= \sum_x \bigg[ -\log \Pf{} (\partial_\tau - \Sigma_x) + \frac{1}{2} \int_0^\beta d\tau_1 d\tau_2 \bigg\{ \Sigma^x(\tau_1, \tau_2) G^x(\tau_1, \tau_2) \\
&- \frac{J_0^2}{4} \left(G^x(\tau_1, \tau_2)\right)^4 - \frac{\tilde{J}^2}{4} \sum_{s>0} s^{-2\alpha} G_x(\tau_1,\tau_2)^2 G_{x+s}(\tau_1, \tau_2)^2  \bigg\}  \bigg] \, . \label{seff}
\end{align}
\subsection{Two point function}\label{s2pt}
Two point functions are obtained by solving the saddle point equations
\begin{align*}
\frac{\delta S_{eff}}{\delta \Sigma(\tau_1,\tau_2)}=0 ~\Rightarrow~&G_x(i\omega) = \frac{1}{-i\omega -\Sigma_x(i\omega)} \, ,\\
\frac{\delta S_{eff}}{\delta G(\tau_1,\tau_2)}=0 ~\Rightarrow~&\Sigma_x(\tau_1,\tau_2) - J_0^2 G_x^3(\tau_1,\tau_2) -\frac{\tilde{J}^2}{2} \sum_{s>0} s^{-2\alpha} G_x(\tau_1, \tau_2) G_{x+s}(\tau_1,\tau_2)^2 \\
&- \frac{\tilde{J}^2}{2} \sum_{s>0} s^{-2\alpha} G_{x-s}^2(\tau_1,\tau_2) G_x(\tau_1,\tau_2) = 0 \, ,\\
\text{or,}~&\Sigma_x(\tau_1,\tau_2) = J_0^2G_x^3(\tau_1,\tau_2) + \frac{\tilde{J}^2}{2} G_x(\tau_1, \tau_2) \sum_{s>0} s^{-2\alpha} \left\{ G_{x+s}(\tau_1,\tau_2)^2 + G_{x-s}(\tau_1,\tau_2)^2  \right\} \, .
\end{align*}
If we assume translational invariance for $G_x$ and $\Sigma_x$, the saddle point equations become
\begin{align}
\nonumber
G^s(i\omega) &= \frac{1}{-i\omega -\Sigma(\omega)} \, ,\\
\Sigma(\tau_1,\tau_2) &= (J_0^2 + \tilde{J}^2 \zeta(2\alpha)) G_s^3(\tau_1,\tau_2) \, .
\end{align}
This is same as the corresponding equation for SYK model with the substitution $J^2 \rightarrow J_0^2 + \zeta(2\alpha) \tilde{J}^2$. For the rest of our paper, we define 
\begin{align}
J_\alpha^2 &:= J_0^2 + \zeta(2\alpha) \tilde{J}^2 \, . \label{Jdef}
\end{align}
Note that in $\alpha \rightarrow \infty$ limit, one has $J_\infty^2 = J_0^2 + \tilde{J}^2$ as expected.

Now we can simply borrow the SYK result and replace $J$ by $J_\alpha$ to get
\begin{align}
G_s(\tau) &= \frac{1}{(4\pi J_\alpha^2)^{1/4}} \frac{\sgn(\tau)}{|\tau|^{1/2}} \, , \label{prop}
\end{align}
for zero temperature and 
\begin{align}
G_s(\tau) &= \frac{1}{(4\pi J_\alpha^2)^{1/4}} \left[ \frac{\pi}{\beta \sin \frac{\pi \tau}{\beta}} \right]^{1/2} \, . \label{finitetemp}
\end{align}
for finite temperature.
For fixed $J_0^2, \tilde{J}^2$, starting with $\alpha = \infty$, as one moves towards smaller values of $\alpha$, $J_\alpha^2$ increases, i.e. more the non-locality stronger the interaction. At $\alpha=1/2$, $J_\alpha^2$ diverges and the two point function vanishes. This signals the inevitability of considering $1/N$ corrections and explicit breaking\footnote{This is in contrast with the SYK situation, where the explicit breaking of reparameterization symmetry is seen only at the level of four point functions.} of reparameterization symmetry. 

For $\alpha < 1/2$, the sum $\sum_{s>0} s^{-2\alpha}$ does not converge, but one can regulate it and take the regulated sum to be $\zeta(2\alpha)$. This regularisation also renders the free energy per fermion finite, thus seems to be a physically sound regularization. Further demanding $J_\alpha^2 >0$ implies $\zeta(2\alpha) > - J_0^2/\tilde{J}^2$. There are infinitely many intervals in $\alpha <1/2$ region satisfying this. There is more to the story though, as we will see in \ref{sstability}.

\subsection{Four point function}\label{s4pt}
Now we compute
\begin{align*}
\frac{1}{N} \mathcal{F}_{xy}(\tau_1,\tau_2;\tau_3,\tau_4) &= \langle G_x(\tau_1,\tau_2) G_y(\tau_3,\tau_4) \rangle - \langle G_x(\tau_1,\tau_2) \rangle \langle G_x(\tau_3,\tau_4) \rangle \, .
\end{align*}
In order to compute this we expand the effective action \refb{seff} about the saddle and then integrate out fluctuations of $\Sigma_x(\tau_1,\tau_2)$ to get a quadratic action for the fluctuations of $G_x(t_1,t_2)$. Then the two point function of fluctuations of $G_x(t_1,t_2)$ leads to 
\begin{align}
\frac{1}{N} \mathcal{F}_{xy} (\tau_1, \tau_2, \tau_3, \tau_4) &= \frac{1}{N} \frac{1}{|G_s(\tau_{12}) G_s(\tau_{34})|} \frac{2}{3J_\alpha^2} \frac{1}{\widetilde{K}^{-1} - S } \, ,
\end{align}
where 
\begin{align}
\nonumber
\widetilde{K}(\tau_1, \tau_2, \tau_3, \tau_4) &= 3J_\alpha^2 G_s(\tau_{13}) |G_s(\tau_{34})| G_s(\tau_{42}) |G_s(\tau_{21})| \, ,\\
\text{and}~S(x,y) &= \delta_{x,y} + \frac{\tilde{J}^2}{3J_\alpha^2} \left( |x-y|^{-2\alpha} (1-\delta_{x,y}) -2\zeta(2\alpha) \delta_{x,y} \right) \, .
\end{align}
To exploit the translational symmetry (which appears only after disorder averaging), it is advisable to go to momentum space. Then one has 
\begin{align}
\frac{1}{N} \mathcal{F}_p (\tau_1, \tau_2, \tau_3, \tau_4) &= \frac{1}{N} \frac{1}{|G_s(\tau_{12}) G_s(\tau_{34})|} \frac{2}{3J_\alpha^2} \frac{1}{\widetilde{K}^{-1} - s(p) \delta(\tau_{13}) \delta(\tau_{24}) } \, ,
\end{align}
where
\begin{align}
s(p) &= 1 + \frac{2\tilde{J}^2}{3J_\alpha^2} \left[ Li_{2\alpha} (\cos p) - \zeta(2\alpha) \right] \, . \label{sp}
\end{align}
Here $Li_{2\alpha}(z)$ is the Polylogarithm function, defined by the following series
\begin{align}
Li_{n}(z) &= \sum_{k =1}^\infty \frac{z^k}{k^n} \, , \label{polylog}
\end{align}
for $|z|<1$. For $n<1$, it has a divergence at $z=1$.

\subsection{Chaos}\label{schaos}
In order to diagnose chaos, one considers the following out of time order correlation function
\begin{align}
F(x,t) &= \frac{1}{N^2} \sum_{j,k=1}^N \langle \chi_{j,x}(t+ \frac{3i\beta}{4}) \chi_{k,0}(\frac{i\beta}{2}) \chi_{j,x}(t+ \frac{i\beta}{4}) \chi_{k,0}(0) \rangle_\beta \, .\label{otoc}
\end{align}
This has the form
\begin{align}
F(x,t) &\sim G(\beta/2)^2 \left( 1 - \frac{\#}{N} e^{\lambda_L t}\right) \, ,
\end{align}	
$\lambda_L$ being the Lyapunov exponent.

The following expression for Lyapunov exponent was derived in \cite{Gu:2016oyy}
\begin{align}
\lambda_L(p) &= \frac{2\pi}{\beta} \left( \frac{3}{2} s(p) - 1/2 \right) \, . \label{lyapu}
\end{align} 
This formula continues to hold in our model, with $s(p)$ given by \refb{sp}. This gives
\begin{align}
\lambda_L(p) &= \frac{2\pi}{\beta}  + \frac{2 \pi  \tilde{J}^2}{\beta J_\alpha^2} \left[ Li_{2\alpha}(\cos p) - \zeta(2\alpha) \right] \, .
\end{align}
Although the Lyapunov spectrum depends on $\alpha$, the maximal Lyapunov exponent $\lambda_L(p=0)$ is independent of $\alpha$ and saturates the chaos bound. This behaviour persists till $\alpha >1/2$.

For $\alpha <1/2$ the situation is quite different. In this regime, $Li_{2\alpha}(p)$ diverges as $p \rightarrow 0$, making $\lambda_L$ unbounded in long wavelength regime. In particular this seems to violate the chaos bound, conjectured in \cite{Maldacena:2015waa}.

One might be hopeful by noting that $J_\alpha^2$ becomes negative across this transition, suggesting the mean field solution is not physically sound. This is not a permanent cure though, since there are infinitely many intervals for $\alpha <1/2$, with $J_\alpha^2 >0$. Chaos bound is violated in all of these intervals. Thus a more permanent solution is needed, which we discuss in \ref{sstability}.

\subsection{Stability} \label{Stability}\label{sstability}
In order to check the stability of a solution, one has to look at one loop free energy. An unstable mode will give imaginary contribution to free energy. In present case the one loop contribution to free energy reads \cite{Maldacena:2016hyu}
\begin{align}
\nonumber
- \beta F \supset &-\frac{1}{2} \int dp \Bigg[ \sum_{m >1} \ln{} [1 -s(p) k(2m)] \\
&+ \int_0^\infty dy \ln{} [1- s(p) k(1/2+iy)] + \sum_n \ln{} [1-s(p)k(2,n)] \Bigg] \, .
\end{align}
From \cite{Maldacena:2016hyu}, one has
\begin{align}
\nonumber
k(h) &= -\frac{3}{2} \frac{\tan \frac{\pi(h - 1/2)}{(h-1/2)}}{h-1/2} \, , \\
\text{and,}~k(2,n) &= 1 -\frac{\sqrt{2} \alpha_K |n| }{\beta J_\alpha} + \dots. \label{ks}
\end{align}
$\alpha_K \sim 2.85$ is a numerical constant.
If the argument of any of these logarithms become negative then the free energy will become imaginary. Let us briefly recollect the situation for SYK model, which can be obtained by putting $s(p)=1$ and removing the momentum integral. In that case, arguments of all logarithms are positive, since $k(2m) <1 , k(2,n) <1$ and $k(1/2+iy) <0$. This ensures the stability of SYK mean field solution.

The stability will continue to persist, if $s(p) \leq 1$, or $Li_{2\alpha}(\cos p) \leq \zeta(2\alpha)~\forall p$. The series expansion \refb{polylog} suggests this is true, since $Li_n(z) \leq Li_n(1) = \zeta(n)$. However, one needs to check the convergence of the series. For $2\alpha<1$ \refb{polylog} ceases to converge at $z=1$, therefore one can not write $Li_n(1) = \zeta(n)$ in this region. In fact since $Li_{2\alpha}(1)$ is divergent and positive (since it is sum of positive numbers), there is always a long wave length regime where $Li_{2\alpha}(\cos p) - \zeta(2\alpha) > 0$ and consequently $s(p) >1$. Thus we conclude the mean field solution \refb{prop} is unstable for $\alpha < 1/2$.

This brings us to the main point of this paper. We see for this model, instabilities stop the system from developing super-maximal chaos. Although this finding is in the context of a particular model, the phenomenon of too much chaos leading to instability is physically quite appealing. Thus we would like to suggest this provides an alternative way to look at the chaos bound, namely $2\pi T$ is the maximal possible value of Lyapunov exponent in a thermal many body quantum system, because any further chaos will render the system unstable.
\subsection{Transport properties and phase structure}\label{sphase}
For long wavelengths, the situation is close to that of SYK model, where we know that the leading contribution to four point function comes from the $h=2$ subspace. Thus in the present case, in order to analyse the long wave length physics, it is advisable to consider the contributions from $h=2$ subspace. Calling this contribution $\mathcal{F}_{p;big}$, following \cite{Maldacena:2016hyu} we have
\begin{align}
3J_\alpha^2 G_s(\tau_{12}) G_s(\tau_{34}) \mathcal{F}_{p;big} &= 2 \sum_{|n| \geq 2} \frac{k(2,n)}{1-s(p) k(2,n)} \Psi_{2,n}(\tau_1, \tau_2) \Psi^*(\tau_3, \tau_4) \, , 
\end{align}
where $\Psi_{2,n}$-s form a basis for $h=2$ eigenspace 
\begin{align}
\nonumber
\Psi_{2,n} (\tau_1, \tau_2) &= \frac{\gamma_n e^{-i \frac{2\pi n}{\beta} \frac{\tau_1 + \tau_2}{2}}}{2 \sin \frac{\pi \tau_{12}}{\beta}} f_n(\tau_{12}) \, , \\
\text{with,}~f_n(\tau)&= \frac{\sin n\pi \tau /\beta}{\tan \pi \tau / \beta}~~, ~~\gamma_n^2 = \frac{3}{\pi^2 |n| (n^2-1)} ~\text{and}~|n| \geq 2 \, ,
\end{align} 
and $k(2,n)$ are the corresponding eigenvalues.\\
In $p \rightarrow 0$ limit one has
\begin{align}
s(p) \sim 1 - \frac{\tilde{J}^2 p^2}{3J_\alpha^2} \zeta(2\alpha -1) \, .
\end{align}
Using this and following steps similar to \cite{Maldacena:2016hyu} one has, in long wave-length limit
\begin{align}
\frac{\mathcal{F}_{p;big}(\tau_1,\tau_2,\tau_3,\tau_4)}{G_s(\tau_{12}) G_s(\tau_{34})} &= \frac{8 \sqrt{2} J_\alpha}{\alpha_K} \sum_{|n| \geq 2} \frac{e^{-\omega_n (\tau_1 + \tau_2 - \tau_3 - \tau_4)/2}}{|n| (n^2-1)} \frac{f_n(\tau_{12}) f_n^*(\tau_{34})}{\omega_n + \frac{\sqrt{2} \pi \tilde{J}^2}{3\alpha_K J_\alpha} \zeta(2\alpha-1) p^2} \, .  \label{fpbig}
\end{align}
Here $\omega_n = \frac{2\pi n}{\beta}$ are the bosonic Matsubara frequencies. We see $\frac{\mathcal{F}_{p;big}(\tau_1,\tau_2,\tau_3,\tau_4)}{G_s(\tau_{12}) G_s(\tau_{34})} $ is Green's function for a diffusion equation with diffusion coefficient
\begin{align}
D_\alpha &= \frac{\sqrt{2} \pi \tilde{J}^2}{3 J_\alpha \alpha_K} \zeta(2\alpha -1) \, . \label{difconst}
\end{align}
In the limit $\alpha \rightarrow \infty$, \refb{difconst}  reproduces the diffusion coefficient of the model considered in \cite{Gu:2016oyy}. As one lowers the value of $\alpha$, the diffusion coefficient increases, but continues to saturate the bound conjectured in \cite{Hartnoll:2014lpa}. This behaviour is analogous to the holographic theories studied in \cite{Blake:2016wvh}, \cite{Blake:2016sud}. The growth of $D_\alpha$ for smaller $\alpha$ is physically expected, since lowering the value of $\alpha$ represents more and more long range interactions.

This growth continues till $\alpha =1$, where $D_\alpha$ diverges. This can not be a physical divergence though, since $s(p)$ is perfectly smooth around $(p=0, \alpha=1)$. 

To understand the situation better, we need to look closely into the function $Li_{2\alpha} (\cos p)$. Since we are interested in small $p$ behaviour, we have to expand $Li_{2\alpha}(z)$ about $z=1$. This expansion is given by
\begin{align}
\nonumber
Li_{2\alpha} (z) &= \Gamma(1-2\alpha) (1-z)^{2\alpha-1} \left[1 + \frac{2\alpha-1}{2} (1-z) + \dots \right]\\
&+ \left[ \zeta(2\alpha) - \zeta(2\alpha-1) (1-z) + \dots \right] \, . \label{polyseries}
\end{align}
For $2\alpha \notin \mathbb{Z}$, \refb{polyseries} is clearly not a Taylor expansion and the fractional powers make it impossible to extend the series for real $z>1$, without compromising the reality of $Li_{2\alpha}(p)$.  Since first fractional power appears as $(1-z)^{2\alpha-1}$,  the leading term in the series is still linear in $p$, as long as $\alpha>1$. Thus we can safely forget the fractional powers for long wavelength physics. 

For $2\alpha \in \mathbb{Z}$, one has to first consider $2\alpha=n+\epsilon$, with $n\in \mathbb{Z}$, and $\epsilon$ small. Then upon taking $\epsilon \rightarrow 0$ limit, one finds log terms. These terms are suppressed for $\alpha >1$.
Finally, for $\alpha =1$ and small $p$, one has
\begin{align}
Li_2 (\cos p) &\sim \frac{p^2}{2} \left( \ln{} \frac{p^2}{2} -1 \right) \, ,
\end{align}
which implies in long wavelength regime 
\begin{align}
\frac{\mathcal{F}^{\alpha=1}_{p;big}(\tau_1,\tau_2,\tau_3,\tau_4)}{G_s(\tau_{12}) G_s(\tau_{34})} &= \frac{8 \sqrt{2} J_{\alpha=1}}{\alpha_K} \sum_{|n| \geq 2} \frac{e^{-\omega_n (\tau_1 + \tau_2 - \tau_3 - \tau_4)/2}}{|n| (n^2-1)} \frac{f_n(\tau_{12}) f_n^*(\tau_{34})}{\omega_n - \frac{\sqrt{2} \pi \tilde{J}^2}{3\alpha_K J_{\alpha=1}} p^2 \left( \ln{} (p^2/2) -1 \right) } \, .  \label{fpbigalpha1}
\end{align}
We see the diffusion coefficient has been logarithmically renormalised:
\begin{align*}
D_1(p) &= -\frac{\sqrt{2} \pi \tilde{J}^2}{3\alpha_K J} \left( \ln{} (p^2/2) -1 \right) \, .
\end{align*} 
Appearance of logarithm also implies that $\frac{\mathcal{F}_{p;big}(\tau_1,\tau_2,\tau_3,\tau_4)}{G_s(\tau_{12}) G_s(\tau_{34})}$ satisfies a differential equation of infinite order, even in long wavelength limit\footnote{This is in contrast with usual situation. Had we kept all powers of momentum, we would have an infinite order differential equation there as well, but in long wavelength limit we can neglect all but few terms to very good approximation. This approximation breaks down at $\alpha=1$.}!  Thus at this value of $\alpha$ the effects of non-locality becomes prominent.

Beyond this point the renormalisation of the diffusion coefficient becomes even stronger and $D(p) p^2$ in the expression of $\mathcal{F}^\alpha_{p;big}$ gets replaced by $- \frac{2 \sqrt{2} \pi \tilde{J}^2}{3 \alpha_K J_\alpha} \Gamma(1-2\alpha) (p^2/2)^{2\alpha-1}$. Formally one can still think of 
\begin{align}
D_\alpha(p) &:= - \frac{2^{5/2-2\alpha} \pi \tilde{J}^2}{3\alpha_K J} \Gamma(1-2\alpha) |p|^{4\alpha-4} \, ,
\end{align}
to be the heavily renormalized diffusion coefficient, which diverges as a fractional power of momentum as $p \rightarrow 0$.
Although diffusion does not seem to be a useful concept anymore, since $\mathcal{F}^\alpha_{big}$ does not satisfy a diffusion equation even approximately. This is rather interesting, since transport seems to be controlled by neither quasi-particles (as in conventional metals) nor diffusion (as in incoherent metals). Since physical nature of the system changes across $\alpha=1$ and thus this should be thought of as a phase transition. 

\paragraph{Thermal Transport:}
In low energy and long wavelength regime, thermal conductance is given by \cite{Gu:2016oyy}
\begin{align}
\kappa'_\alpha(\omega,p) &= \frac{N c_v (\alpha) D_\alpha \omega^2}{\omega^2 + (D_\alpha p^2)^2} \, , 
\end{align}
where $c_v (\alpha)= \frac{\pi \alpha_K}{16 \sqrt{2} \beta J_\alpha}$ is the specific heat. For the constant mode, conductance is independent of frequency and is given by $\kappa'_\alpha = N c_v D_\alpha$. As one approaches $\alpha \rightarrow 1_+$, this diverges logarithmically in momentum. Beyond this, $\kappa'_\alpha$ diverges as a fractional power of momentum. 


\paragraph{Butterfly velocity:}
In order to discuss butterfly velocity, one first considers the following out of time order correlation function \refb{otoc}, which has the form $G(\beta/2)^2(1-\mathcal{F}/N)$. One has \cite{Gu:2016oyy} 
\begin{align}
\frac{\mathcal{F}_p(t)}{G(\beta/2)^2} \sim - \frac{1}{b(p)} e^{\frac{2\pi}{\beta} \left[ 1- 3b(p) \right]t} \, ,~\text{where}~~b(p)= \frac{\sqrt{2} \alpha_K}{4\pi J} \left( \frac{2\pi}{\beta} + Dp^2 \right)
\end{align}
This has a pole at $b(p)=0 \Rightarrow p = \pm i \frac{2\pi}{D \beta}$. When transformed back to position space, the contour integral picks contribution from the relevant pole. This gives
\begin{align}
\mathcal{F}(x,t) \sim e^{\frac{2\pi}{\beta} (t - x/v_B)}\, , ~~~v_B^2= \frac{2\pi D}{\beta} \, .\label{butterfly}
\end{align}
$v_B$ is called butterfly velocity, the velocity at which butterfly effect spreads in space \cite{Shenker:2013pqa}. 

\refb{butterfly} remains valid for $\infty>\alpha >1$ with $D$ in \refb{butterfly} replaced by $D_\alpha$. Starting with $\alpha \rightarrow \infty$, as one lowers $\alpha$, butterfly velocity increases. This is expected since interaction is more and more long range. As $\alpha \rightarrow 1$, the butterfly velocity diverges.

For $\alpha \leq 1$, the dynamics is no more diffusive thus \refb{butterfly} is not expected to hold any more\footnote{A mathematical way to see this is the following. For $\alpha<1$, $D_\alpha$ being a non trivial function of momentum, $b(p)$ does not have any simple zero.}. It would be interesting to understand what plays the role of butterfly velocity in this phase.


\paragraph{Ergodic dynamics}
For a metal wire of length $L$, ergodic dynamics takes over \cite{Altland:2017eao} for time scales longer than the ergodic time $t_{erg}=L^2/D$, $D$ being the diffusion coefficient. For shorter time scales the dynamics is diffusive and is primarily governed by the the eigenmodes of the diffusion operator $D \partial_x^2$. Note that stronger the diffusion, shorter is the diffusive time window and sooner is the onset of ergodicity. For dirty metals $t_{erg}$ is called ``Thouless time". Since we have a disordered wire, we will henceforth refer $t_{erg}$ as Thouless time.

In the present case, $D_\alpha \sim \zeta(2\alpha -1)$. As one starts from $\alpha \rightarrow \infty$ and moves towards smaller $\alpha$, $D_\alpha$ increases. However in order to discuss $t_{erg}$, we need to regulate the length of the system, which is naively infinite in the present case. After appropriate regularisation (see \ref{reg}), one finds the regularized length of the system $L_\alpha$ increases faster than $D_\alpha$ with decreasing $\alpha$. Consequently the diffusive time window broadens and at $\alpha \rightarrow 1$ dynamics remains diffusive for ever. Ergodicity never sets in.

For $\alpha < 1$, the dynamics ceases to be diffusive and $D_\alpha$ does not seem to be very useful concept. It is not clear what plays the role of $t_{erg}$ in this phase.
\paragraph{A stranger metal:} In a strange metal, i.e. system without quasiparticles, it has been argued that Lyapunov time $\tau_L= 1/\lambda_L$ and butterfly velocity $v_b$ can play the roles of quasiparticle mean free time and Fermi velocity respectively, which are crucial for transport properties of a normal metal. This suggests that transport in such systems is related to chaos. In fact it has been proposed \cite{Blake:2016wvh}, \cite{Blake:2016sud}, \cite{Blake:2017qgd}  that thermal conductivity in such systems obey $D \sim v_b^2 \tau_L$.

The phase we encounter for $1/2<\alpha<1$, is rather strange from this perspective. This is a phase without quasiparticles and maximally chaotic. However due to peculiar momentum dependence of density-density correlators, diffusion does not seem to be a useful concept in this regime. Although a phase without quasiparticles, this seems to be qualitatively different from usually discussed strange metals, where diffusion is a meaningful concept. We call this phase stranger metal.
\section{Discussion} \label{sdscsn}
In this paper, we have attempted to understand the chaos bound \cite{Maldacena:2015waa} from a slightly different angle, namely by asking - ``what goes wrong if a system intends to develop super-maximal chaos"? To gather intuition about possible answer, we study a SYK lattice model, with a tuneable parameter, such that the Lyapunov spectrum varies as the parameter is varied. The largest Lyapunov exponent remains constant though (and equals its maximal value $2\pi T$) until this parameter touches $1/2$ (from right), where the system violates the chaos bound, apparently jeopardising the conjecture of \cite{Maldacena:2015waa}. However one closer inspection, one discovers the mean field solution becomes unstable precisely at this point, thus salvaging the chaos bound conjecture \cite{Maldacena:2015waa}. 

We suspect that the phenomenon of a system developing instabilities while trying to violate the chaos bound, may be of general significance. In fact this suggests the question we posed in the beginning, may have a simple answer- ``super-maximal chaos leads to instability". One wonders if it is possible to develop similar understanding of conjectured bounds on diffusion coefficient \cite{Hartnoll:2014lpa} and viscosity \cite{Kovtun:2004de}?

Our analysis is performed in large $N$ limit. Although we do not expect $1/N$ corrections to change the general lessons, it would be advisable to chek this explicitly.

As a byproduct of our analysis of this model, we discovered a novel phase that appears before instability hits the system. We called this phase ``stranger metal". This is a phase sans quasiparticles, just like strange metals, but is not diffusive due to peculiar momentum dependence of density density correlators. Again, it would be interesting to check how $1/N$ corrections affect this phase. Another natural question is whether long range interaction (which is the present case) is crucial for such a phase. In other words it would be interesting to investigate if some model with local interactions exhibits similar phase. 

Some other questions to explore could be:  what is the nature of thermalization and spectral properties of the system? How do they change across the phase transition\footnote{These questions for SYK model and its tensorial cousins have been studied in \cite{Zhang:2017jvh,Garcia-Garcia:2017pzl,Sonner:2017hxc,Krishnan:2016bvg,Hunter-Jones:2017raw}.}? How is the phase structure modified when various perturbations are turned on? Is it possible to propose an experimental realisation for this model\footnote{For original SYK model experimental realisation has been proposed in \cite{chew}}?

\vspace{1cm} \noindent {\bf Acknowledgements:} This work was conducted within the ILP LABEX (ANR-10-LABX-63) supported by French state funds managed by the ANR within the Investissements d'Avenir program (ANR-11-IDEX-0004-02) and supported partly by the CEFIPRA grant 5204-4.

\newpage
\appendix
\section{Regulariziation of Thouless Energy} \label{reg}
Most natural way to regularise the system seems to be to define the system on a circle. We compactify the system on a circle with $2L+1$ sites, which are labelled as $0,1, \dots, 2L$. The distance between two sites $x$ and $y$, can either be taken to be $|x-y|$ or $2L+1-|x-y|$. We take the interaction strength between two such sites to be $|x-y|^{-\alpha} + (2L+1-|x-y|)^{-\alpha}$. Very large $L$, one can approximate
\begin{align}
\nonumber
|x-y|^{-\alpha} + (2L+1-|x-y|)^{-\alpha} &\sim |x-y|^{-\alpha} ~~~\text{for} |x-y| \leq L \\
&\sim (2L+1-|x-y|)^{-\alpha}  ~~~\text{for} |x-y| > L \, . \label{approx}
\end{align}
This approximation fails terms for which $|x-y| \sim L$. This is not a problem though since the contribution of such terms to any physical quantity is of order $L^{-\alpha}$, which itself is vry small for large $L$. \ref{approx} in turn leads to the following approximation 
\begin{align}
\sum_{y \neq x; \atop y=0}^{2L} \left( |x-y|^{-\alpha} + (2L+1-|x-y|)^{-\alpha}  \right)^2 &\sim 2\sum_{x-y=1}^{L} |x-y|^{-2\alpha}  \sim \zeta(2\alpha) \, . \label{approx2}
\end{align}
There are two conditions for \refb{approx2} to work well: a) $L^{-\alpha}$ should be very small for \refb{approx} to hold\footnote{Even then \refb{approx} is not a good approximation for $|x-y| \sim L$, since the term kept is nearly same as the term thrown. But this does not affect the physics since such terms are anyway very small.} and b) $\sum_{k=1}^L k^{-2\alpha}$ should be quite close to $\zeta(2\alpha)$. 

In order to be able to discuss Thouless time, which involves the diffusion coefficient $D \sim \zeta(2\alpha-1)$, we need a little more: $\sum_{k=1}^L k^{-2\alpha+1}$ should well approximate $\zeta(2\alpha-1)$ as well. This is hardly an extra criteria for large or even moderate $2\alpha$, but becomes crucial when $2\alpha - 1 \rightarrow 1_+$. 

We choose the length $L_\alpha$ (we have put the subscript $\alpha$ to stress that the regularised length depends on $\alpha$) such that
\begin{align}
\sum_{k=L_\alpha}^{\infty} k^{-(2\alpha-1)} < \epsilon \zeta (2\alpha-1) \, ,
\end{align}
where $\epsilon$ is some number smaller than $\zeta(2\alpha -1)^{-1}$. $L_\alpha$ can be approximated by
\begin{align*}
\int_{L_\alpha}^\infty dx x^{1-2\alpha} &\sim \epsilon \zeta (2\alpha-1) \\
\frac{L_\alpha^{2-2\alpha}}{2\alpha -2} &\sim \epsilon \zeta (2\alpha-1) \\
L_\alpha^{2-2\alpha} &\sim \epsilon (2\alpha -2) \zeta (2\alpha-1) \sim \epsilon\\
L_\alpha &\sim \epsilon^{1/(2-2\alpha)} \, .
\end{align*}
Thus $L_\alpha$ diverges exponentially as $\alpha \rightarrow 1_+$. This is much faster than the divergence of $D$ which is only polynomial. Therefore regularised Thouless energy $E_c = D/L_\alpha^2$ vanishes and regularised Thouless time diverges, as $\alpha \rightarrow 1_+$.
 


\end{document}